\documentclass[a4paper,11pt]{article}

\usepackage{jinstpub} 

\title{\boldmath Precise Mirror Alignment and Basic Performance of the RICH Detector of the NA62 Experiment at CERN}

\author[a,b]{G.~Anzivino,}
\author[b]{M.~Barbanera,}
\author[c,d]{A.~Bizzeti,}
\author[a,b]{F.~Brizioli,}
\author[d]{F.~Bucci,}
\author[d,e]{A.~Cassese,}
\author[b]{P.~Cenci,}
\author[b]{B.~Checcucci,}
\author[d]{R.~Ciaranfi,}
\author[b,f,1]{V.~Duk \note{Corresponding author.},}
\author[g]{J.~Engelfried, }
\author[g]{N.~Estrada-Tristan, }
\author[d,e]{E.~Iacopini,}
\author[a,b]{E.~Imbergamo,}
\author[d,e]{G.~Latino,}
\author[d,e]{M.~Lenti,}
\author[a,b]{R.~Lollini,}
\author[b]{M.~Pepe,}
\author[b]{M.~Piccini,}
\author[d,e]{R.~Volpe}

\affiliation[a]{Dipartimento di Fisica e Geologia dell'Universit\`a di Perugia, \\
Via A. Pascoli, Perugia, Italy}
\affiliation[b]{INFN -- Sezione di Perugia, \\
Via A. Pascoli, Perugia, Italy}
\affiliation[c]{Dipartimento di Scienze Fisiche, Informatiche e Matematiche
   dell'Universit\`a di Modena e Reggio Emilia, \\
Via G. Campi, Modena, Italy}
\affiliation[d]{INFN -- Sezione di Firenze, \\
Via G. Sansone, Sesto Fiorentino, Italy}
\affiliation[e]{Dipartimento di Fisica e Astronomia dell'Universit\`a  di Firenze, \\
Via G. Sansone, Sesto Fiorentino, Italy}
\affiliation[f]{School of Physics and Astronomy, University of Birmingham, \\
Edgbaston Park Rd., Birmingham, United Kingdom $^2$
  \note{Funded by the EU Horizon 2020 research and innovation programme 
(Marie Sklodowska-Curie grant No 701386).}}
\affiliation[g]{Instituto de F\'isica, Universidad Aut\'onoma de San Luis Potos\'i, \\
Alvaro Obreg\'on, Zona Centro, San Luis Potos\'{\i}, Mexico$^3$
\note{Funded by Consejo Nacional de Ciencia y Tecnolog\'ia (CONACyT) and Fondo de Apoyo a la Investigaci\'on (UASLP).} }

\emailAdd{Viacheslav.Duk@cern.ch}

\abstract{
The Ring Imaging Cherenkov detector is crucial
for the identification
of charged particles
in the NA62 experiment at the CERN SPS.
The detector commissioning was completed in 2016
by the precise alignment of mirrors using
reconstructed tracks. 
The alignment procedure
and measurement of the basic performance
are described.
Ring radius resolution,
ring centre resolution, single hit resolution and mean number of hits per ring
are evaluated for positron tracks.
The contribution of the residual mirror misalignment to the performance is calculated.

}


\begin{document}
\maketitle
\flushbottom

\section{Introduction}
\label{Intro}
\begin{figure}[h]
\centering
\includegraphics[width=13cm]{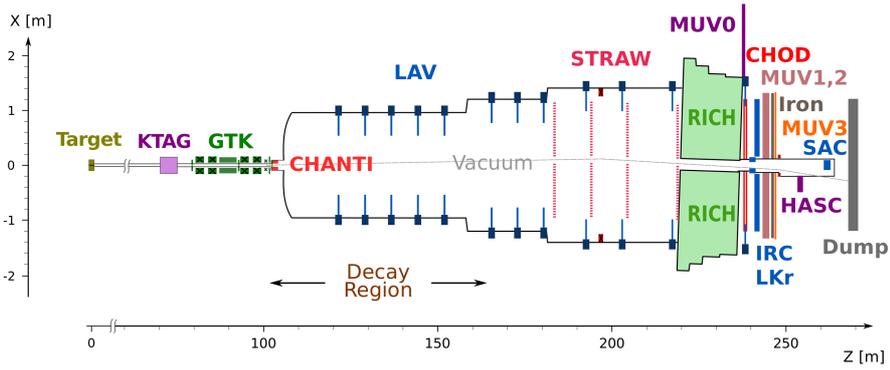}
\caption{NA62 experimental setup. The beam goes in the positive Z direction. 
The positive direction of the Y axis is vertical.}
\label{layout}       
\end{figure}

\hspace{15pt}
The NA62 experiment at CERN is aimed at measuring  the ultra 
rare decay $K^+\to\pi^+\nu\bar\nu$ (BR$\sim$10$^{-10}$).
The BR measurement with 10$\%$ precision will allow
to probe New Physics at mass scales up to O(100)~TeV.
The experimental setup is shown in figure~\ref{layout} and
 described in detail in~\cite{na62_detector_paper}.
A 400 GeV/c proton beam from the CERN SPS  impinging on a Beryllium target produces a 750~MHz hadron beam
of 75 GeV/c with $\sim$6$\%$ of K$^+$ particles.
Kaons are identified by the KTAG detector, a differential Cherenkov counter.
The momentum of beam particles is measured by the beam tracker (GTK).
The momentum of secondary particles is measured by 
a magnetic spectrometer with
Straw chambers (STRAW) operating in vacuum.
The system of hodoscope counters (CHOD)
consisting of scintillator slabs and tiles 
measures the track crossing time and contributes
to the L0 trigger, as well as the 
Ring Imaging CHerenkov detector (RICH).
The iron/scintillator calorimeters (MUV1,2)
identify pions and muons, while the electron/positron identification (ID)
is performed by the electromagnetic calorimeter filled with 
Liquid Krypton (LKr).
A fast muon veto (MUV3)
identifies muons and provides
L0 trigger signals. 
The photon veto system 
covers the angular range up to 50 mrad and
includes four detectors: LAV, LKr, 
 IRC and SAC.
The CHANTI detector placed after the third station of GTK
identifies upstream inelastic interactions and muon halo.
Additional veto detectors MUV0 and HASC
are used to detect pions from the  $K^+\to\pi^+\pi^-\pi^+$ decay
escaping from the STRAW acceptance.

\section{RICH detector}
\hspace{15pt}
One of the main backgrounds to the $K^+\to\pi^+\nu\bar\nu$ decay comes from 
  $K^+\to\mu^+\nu_\mu$ which is suppressed by applying specific selection 
criteria on kinematic variables and making use of the different stopping power of muons and pions 
in MUV1 and MUV2. The RICH detector is needed to further reject the muon contamination in 
the pion sample by a factor of at least 100 in the momentum range between 15 and 35 GeV/c. 
The upper bound of this range is driven by the kinematic suppression of the other 
principal background, the $K^+\to\pi^+\pi^0$ (K2$\pi$)
decay. To distinguish between muons and pions at 35 GeV/c, the RICH should have a Cherenkov threshold for pions
around 12--13 GeV/c which means that the full efficiency of the RICH is achieved at 15~GeV/c. 
The choice of this lower bound is also favoured by studies of other backgrounds.

\begin{figure}[b]
\centering
\includegraphics[width=13cm]{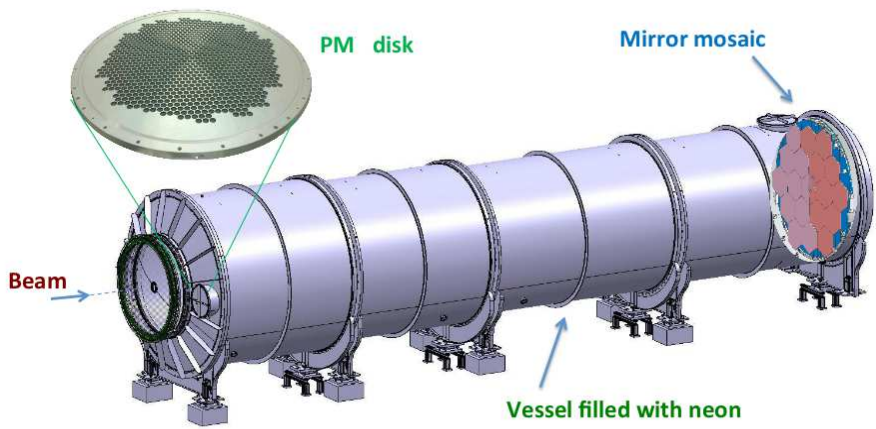}
\caption{RICH detector layout. 
The zoom is
done for one of two photomultiplier disks. The mirror mosaic is made visible on the right.
The right part of mirrors (shown in dark pink) reflects light towards the zoomed disk, while the other half of the mosaic 
(shown in light pink)
is oriented towards the second disk (not~seen).
}
\label{vessel}   
\end{figure}

\begin{figure}[h]
\begin{minipage}[t]{0.45\textwidth}
\centering
\includegraphics[width=6.0cm , angle=0]{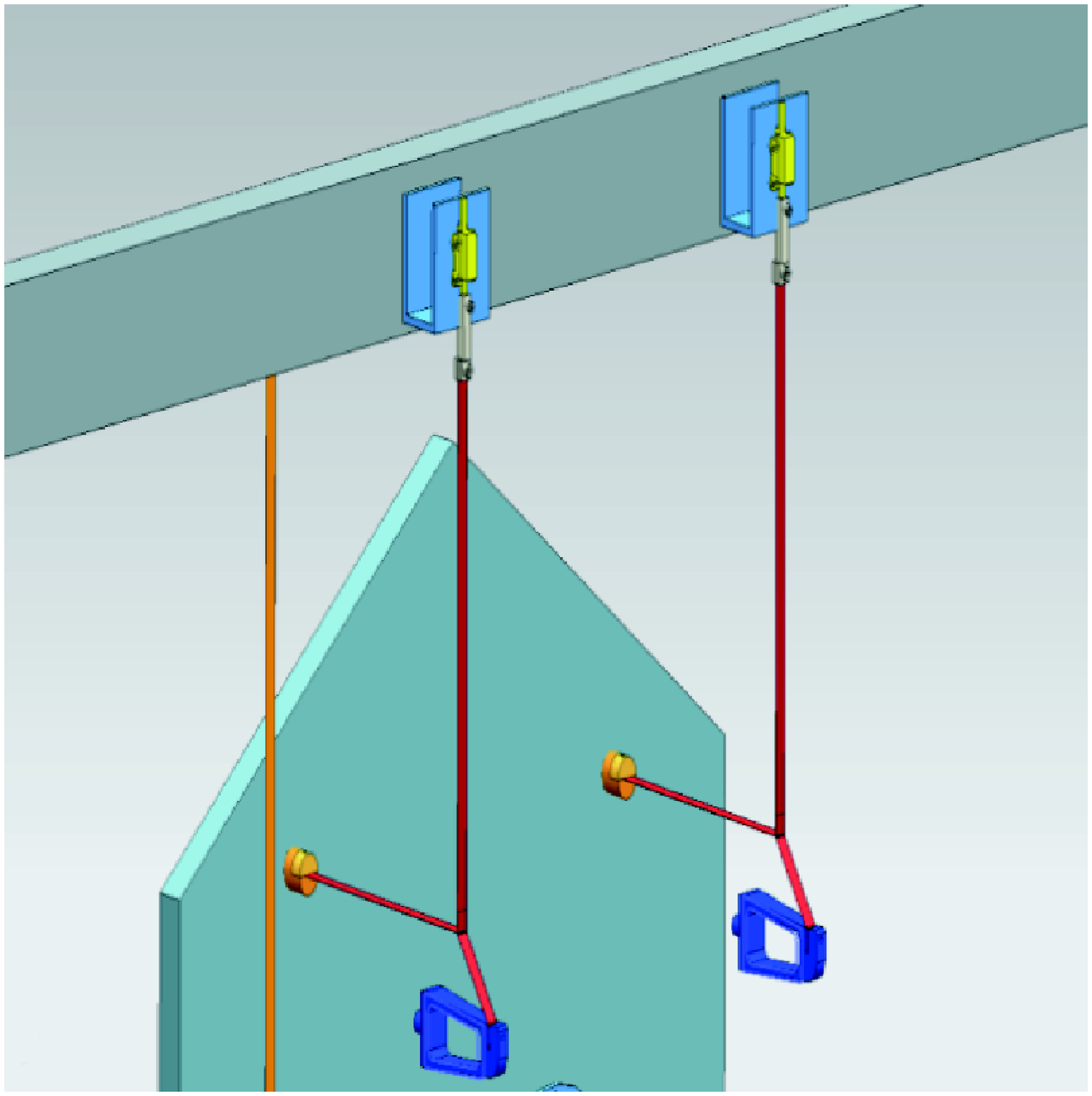}
\end{minipage}  
\hspace{0.2cm}
\begin{minipage}[t]{0.45\textwidth}
\centering
\includegraphics[width=6.55cm , angle=0]{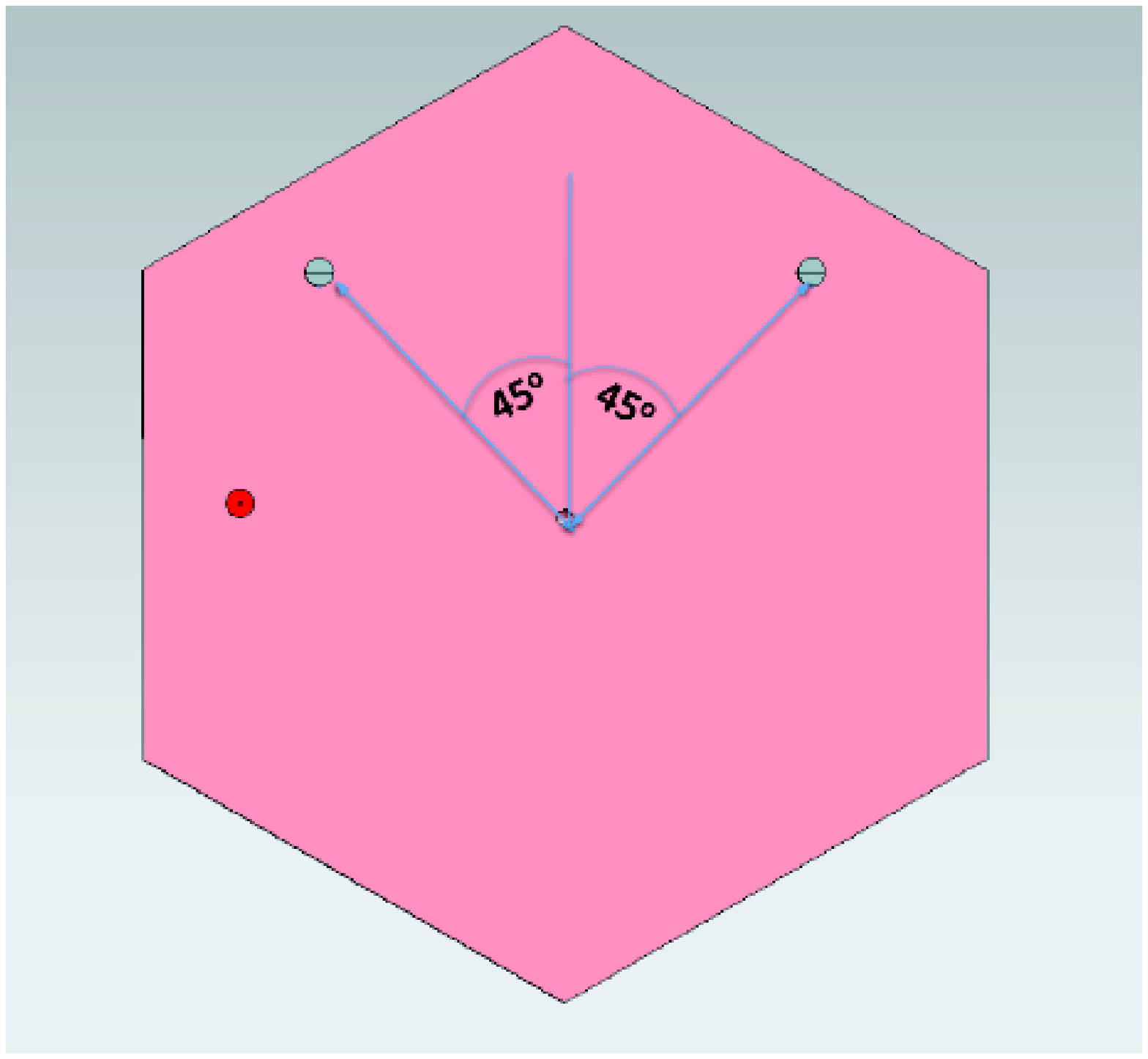}
\end{minipage} 
\caption{Arrangement of the mirror orientation
system, view from the downstream part of the setup. The anti-rotating ribbon
is connected to the mirror at point 3.
Two stabilizing ribbons are connected to piezo motors 
L and R and to the mirror 
at points 1 and 2 respectively
via the transmission tool. $R_{con}$
is the distance between the barycentre O and the ribbon connection points.}
\label{mirrors}
\begin{picture}(1,1)
\put(90,205){L}
\put(140,220){R}
\put(55,118){1}
\put(105,133){2}
\put(45,80){3}
\put(258,212){1}
\put(335,212){2}
\put(245,175){3}
\put(296,152){O}
\put(265,182){\tiny $R_{con}$}
\put(325,182){\tiny $R_{con}$}
\end{picture} 
\end{figure}

The RICH detector is shown in figure~\ref{vessel}.
The core part of the detector is the mirror system~\cite{mirror_system_paper}.
It 
consists of 18 hexagonal (350 mm side) 
and two semi-hexagonal mirrors which are 
placed in the central part and
cut to accomodate the beam pipe.
The focal length of all mirrors is $f$=17~m. 
The mirror orientation is provided by two stabilizing aluminium ribbons
connected to the mirror at one end (at a distance of $R_{con}\sim$250 mm
from the barycentre) and to a piezo motor at the other end
via the transmission tool.
A third anti-rotating ribbon prevents the mirror rotation around the longitudinal axis.
The ribbon arrangement is shown in figure~\ref{mirrors}.
Piezo motors move ribbons with 1 nm step.

\begin{figure}[h]
\centering
\includegraphics[width=8.5cm]{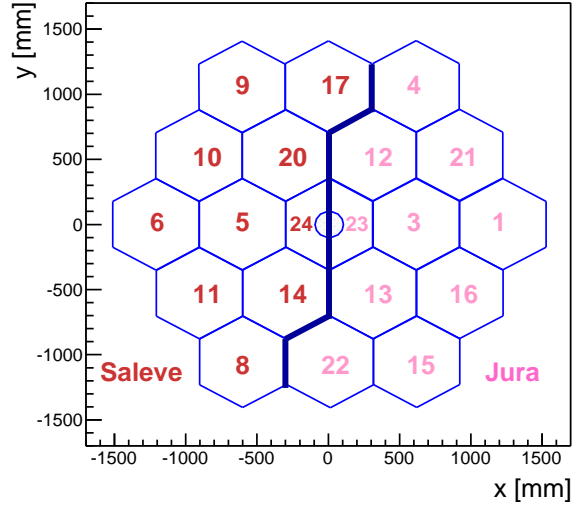}
\caption{RICH mirror numbering as seen from the downstream part of the setup. 
The axes direction is the same as for the NA62 reference frame in figure~\ref{layout}.
Jura and Saleve groups are separated by the dark blue line.
For the definition of Jura and Saleve, see section 2.
}
\label{mirror_numbering}   
\end{figure}

To avoid the loss of the reflected light
interacting
 with the beam pipe, the mirrors
are divided in two groups referred to as {\bf Jura} and {\bf Saleve}
 with centres of curvature of mirror surface, respectively,
to the right and to the left 
of the beam pipe, as seen from the downstream part of the setup. 
Figure~\ref{mirror_numbering} illustrates the mirror numbering and Jura--Saleve orientation.
The Jura group is shown in light pink, the Saleve one is in dark pink,
the same colors are used in figure~\ref{vessel}.

Two photomultiplier (PM) disks
are placed in the focal plane of each mirror orientation group and
are located
at about 1.5 m to the left and to the right of the beam pipe, outside 
the area illuminated by charged particles from kaon decays in the fiducial volume.
Each disk contains 976 PMs.
The PM disk diameter is $\sim$600 mm. 
To enhance light collection,
 Winston  cones~\cite{winston_cone_paper} with 
the outer diameter d$_{cone}$=18 mm are carved  in  the  disks  and  covered  with 
aluminized Mylar (one cone per PM).
The inner diameter of Winston cones is equal to 
the diameter of
 the PM sensitive area  d$_{PM}$=7.5 mm.

\section{Precise mirror alignment}

\subsection{Alignment procedure}
\label{alignment_procedure}
\hspace{15pt}
The best performance of the RICH detector is achieved when the mirrors are aligned
with the highest possible precision.
During the installation a preliminary laser alignment was performed 
for each mirror with the accuracy of $\sim$500 $\mu$rad
in terms of mirror orientation~\cite{mirror_system_paper}.
For a more precise alignment using reconstructed tracks
a dedicated procedure has been developed.
For each orientation group (Jura or Saleve, see figure~\ref{mirror_numbering})
 a reference mirror was chosen and all other mirrors were 
aligned with respect to that mirror. A natural choice for the reference mirror is a semihexagonal one,
for which the remotely controlled rotation is limited to one degree of freedom,
i.e. only one ribbon can be moved  using piezo motors.

The fine alignment procedure consists of three steps. 
It starts from the measurement
of the {\bf absolute misalignment} (i.e. with respect to the nominal orientation)
of all 20 mirrors. Events with one track in the STRAW  and one RICH ring candidate are selected for the analysis.
The RICH ring is required to be completely within the PM acceptance.
A circle at the mirror plane
centered on the track impact point
and having the same radius as the ring
is required to be 
 within a single mirror
(``single mirror `` condition).
The absolute misalignment of a mirror
is the mean value of the difference between the real
ring centre position from the ring fit and the expected position.
The latter corresponds 
 to the nominal mirror orientation
and is obtained by extrapolating
the track to the PM plane as if it were a photon with the direction of the track
reflected by a mirror with the nominal centre of curvature.

At the second step, the {\bf  relative misalignment}  
of each hexagonal mirror
 is calculated. 
The relative misalignment is defined as the difference
in the absolute misalignment
between a mirror and
the reference mirror of a corresponding group.
Using a simple model with the ideal ribbon geometry~\cite{mirror_system_paper},
 it is linearly translated
to the piezo motor movement needed to compensate the relative misalignment: 
\begin{equation}
\Delta l_L = \frac{R_{con}}{2\sqrt{2}f} (-X_{rel} + Y_{rel})~~ ; ~~~~~~
\Delta l_R = \frac{R_{con}}{2\sqrt{2}f} (X_{rel} +  Y_{rel}).
\label{formula} 
\end{equation}
Here $X_{rel}$ and $Y_{rel}$ are the relative misalignment values,
$\Delta l_L$ and $\Delta l_R$ are the movements 
of piezo motors L and R
needed to compensate this misalignment,
 $R_{con}$ is the distance
between the ribbon connection point to the mirror and 
the mirror barycentre (for the definition of L, R and  $R_{con}$ see figure~\ref{mirrors}),
 $f$ is the mirror focal length. 
Each mirror
is rotated by moving two piezo motors according to the calculated values $\Delta l_L$ and $\Delta l_R$. 
After the first movement
the misalignment is measured again, and
the change in the relative misalignment is translated back to the effective
movement of piezo motors
$\Delta l_{L, eff}$ and $\Delta l_{R, eff}$ 
(i.e. the piezo motor movement which would produce
that change in the misalignment in case of the ideal ribbon geometry).
For each piezo motor a calibration constant is calculated 
by comparing the effective
and real movement: 
c$_L = \Delta l_L / \Delta l_{L, eff}$ for a left motor,
 c$_R = \Delta l_R / \Delta l_{R, eff}$ for a right one.
Further piezo movements are performed taking into account
these constants,
i.e. $\Delta l_L$ and $\Delta l_R$ calculated from~(\ref{formula})
are multiplied by c$_L$ and c$_R$ respectively.

The final step of the procedure is the calculation of {\bf global offsets} and the {\bf residual misalignment}.
A global offset is the average absolute misalignment of a group of mirrors with the same centre of curvature (Jura or Saleve).
To calculate a global offset,
events with hits in a single PM disk (and hence only one group of mirrors illuminated)
 are selected 
(``single mirror `` condition is not applied),
and the absolute misalignment is measured. The difference between the absolute misalignment of a mirror
and the global offset is referred to as the residual misalignment.

 For rings with photons from a single group of mirrors the performance does not depend 
on how the global offset of that group is defined, while
 for rings
where mirrors of both groups are illuminated
such definition of global offsets 
provides the minimal average spread of hit coordinates due to the 
residual misalignment
and hence the best single hit resolution.
A simpler alternative could be to define a global offset as the absolute misalignment of the reference mirror
(the residual misalignment in this case would be equal to the relative one), but in this case
the best performance 
will be achieved only for rings with hits from reference semihexagonal mirrors, while for rings
with hits from hexagonal Jura and Saleve mirrors (for example,  $\#$13--14 or  $\#$12--20
 in figure~\ref{mirror_numbering})
a larger relative misalignment will take place, that will result in a worse single hit resolution.

The procedure of piezo motor movement is repeated iteratively
until the final accuracy is achieved: $\pm$1 mm in terms of the relative misalignment,
 or $\sim$30 $\mu$rad in terms
of the mirror angular orientation.
The latter number comes from the relation 
 $\Delta \theta = \Delta r / 2f$,
where $\Delta \theta$ is the mirror rotation, $ \Delta$r is the corresponding movement
of the ring centre in the focal plane and $f=17$~m is the mirror focal length.
At each iteration step the global offsets calculated at the previous step
are used as initial global offsets.

Global offsets and residual misalignment values are stored in a metadata file and used in the analysis chain.
At the RICH reconstruction level (when only the RICH hits are used,
no track information is available)
global offsets are subtracted from hit coordinates before the standalone ring fit is performed.
At the analysis level the ring fit can be improved by using the track information.
In this case the absolute misalignment 
(i.e. the sum of the global offset and the residual misalignment) 
of the track-pointed mirror 
 is subtracted from the hit coordinates before the track-seeded ring fit.
Such offset subtraction is driven by the fact that 
 the main part of photons is reflected by the mirror where the track points.
This can be easily obtained from 
geometrical considerations assuming a uniform spatial density of photons in the mirror plane.
Moreover, due to the properties
of the Cherenkov radiation (dN/dz=const and hence dN/dr=const,
where r is the radial coordinate with respect to the track impact point 
of a photon emitted at z)
this density is proportional to 1/r,
 which leads to even higher photon concentration
around the impact point.

\subsection{Alignment in 2016}
\label{alignment_in_2016}
\begin{figure}[h]
\centering
\includegraphics[width=15cm]{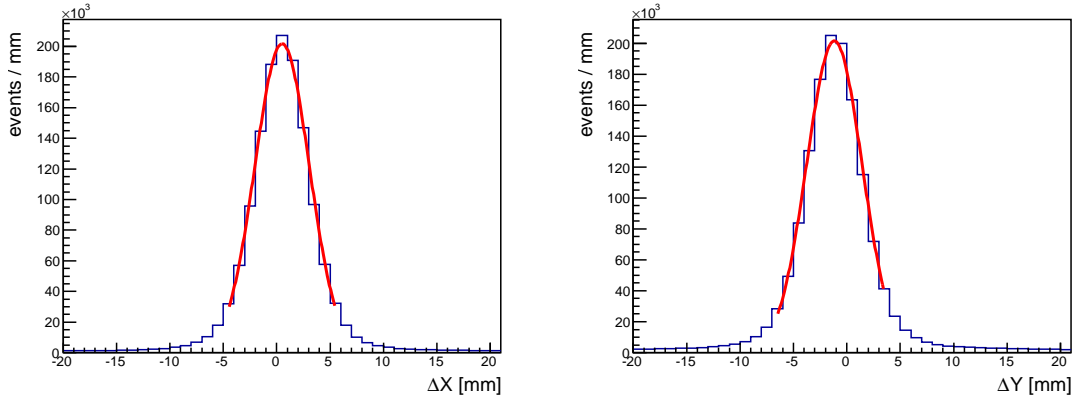}
\caption{Alignment of mirror $\#$5 (first step of the procedure, see~\ref{alignment_procedure}).
 $\Delta$X and $\Delta$Y are the differences between 
the measured and expected
ring centre coordinate.
Initial global offsets are subtracted.
The distributions are fitted with a gaussian. 
The absolute misalignment is the sum of the initial
global offset and the gaussian mean value.
The gaussian width is $\sigma \sim$2.7 mm.}
\label{misalignment}   
\end{figure}

\hspace{15pt}
In 2016 the alignment procedure was fully accomplished for the first time. 
A typical measurement of the absolute misalignment
is shown in 
figure~\ref{misalignment}. 
The accuracy of the misalignment measurement is estimated to be 0.1 mm, 
the main contribution coming from the fitting procedure.
Two contributions determine the  width of the  $\Delta$X and $\Delta$Y distribution:
the uncertainty of ring centre and spectrometer resolution.
 The latter is small 
and can be estimated by multiplying 
the mirror focal length by
the STRAW angular resolution $\sigma_{\theta x}$ or $\sigma_{\theta y}$.
In the assumption that $\sigma_{\theta x} \approx \sigma_{\theta y} = \sigma_\theta/\sqrt{2}$,
where the value $\sigma_\theta$ is taken from~\cite{na62_detector_paper},
the spectrometer contribution to the widths does not
exceed 0.6 mm.

\begin{figure}[h]
\centering
\includegraphics[width=15cm]{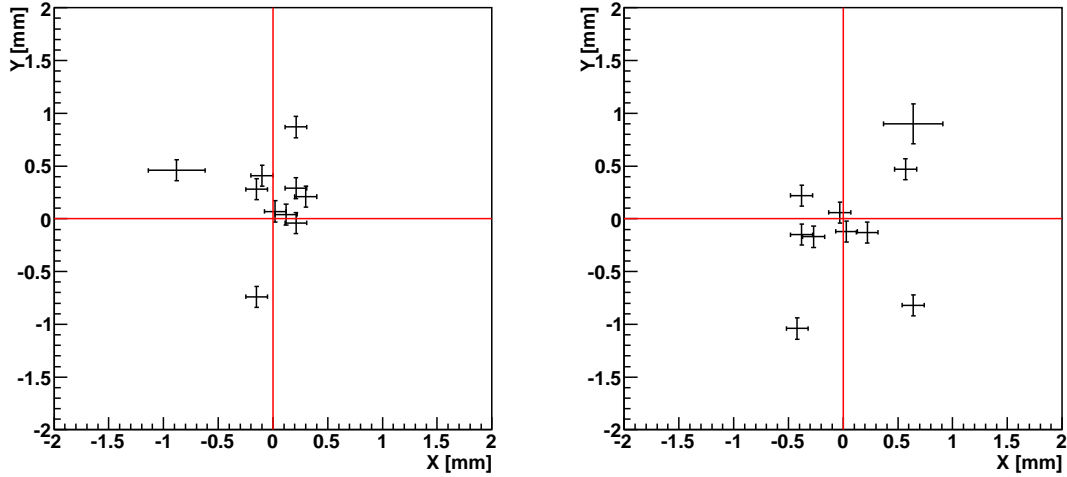}
\caption{Final results of the RICH mirror alignment. Residual misalignment values X and Y
 are shown
for Jura (left) and Saleve (right) mirror groups. Each point corresponds to one mirror.
For the definition of Jura and Saleve, see section 2.
}
\label{alignment}  
\end{figure}

The global offsets (X$_{global}$, Y$_{global}$) at the end of the alignment procedure 
were equal to (20.0,~20.1) mm for Jura and (20.1,~9.5) mm for Saleve.
The final results of the residual misalignment measurement are shown in figure~\ref{alignment}.
 The precision of the overall procedure
 is $\sim$1 mm and is limited by hysteresis effects in the ribbon-mirror system:
for small movements there is no longer linearity between piezo motor 
and ring centre movement, so the iterative procedure does not necessarily converge.
The values of the residual misalignment are given in appendix (table~\ref{table_misalignment}).

In 2017 the mirror alignment was monitored on a monthly basis and remained
stable during the data taking period.

\section{Basic performance in 2016}
\hspace{15pt}
The RICH detector was designed to provide the muon suppression
at the level of O(100) in the pion sample 
and measure the downstream time with O(100) ps precision.
The corresponding performance characteristics (i.e. pion ID efficiency,
muon mis-ID probability, event time resolution) depend on the event selection
and their measurement is beyond the scope of this paper. The preliminary results 
are reported 
in~\cite{na62_detector_paper}. 
Apart from the event selection, these characteristics are determined by more
fundamental performance parameters like single hit resolution and the average
number of hits per event which are traditionally evaluated for electron/positron tracks
in order to avoid the momentum dependence.

In this section the measurement of the basic performance
of the RICH detector is described 
which has been performed on rings fully contained in the detector acceptance 
(to avoid edge effects)
and includes 
the following parameters:
 ring radius resolution,
 ring centre resolution,
 single hit resolution and
mean number of hits per ring.

\subsection{Event selection}
\hspace{15pt}
The positron sample has been collected by the tight selection of the 
$K^+\to e^+ \nu_e \pi^0$ (Ke3) decay events. The selection criteria can be grouped
into four categories: 
 one track selection,
 particle ID,
 kinematics and
 RICH selection.

The {\bf one track selection} 
requires one track events with 
a track including hits from all chambers and lying in the acceptance
of each STRAW station, LKr, CHOD and MUV3. 
Other track requirements are: time within $\pm$10 ns from the trigger time, $\chi^2$
less than 20, momentum between 12 and 40~GeV/c.

The {\bf positron ID} is based on the information from calorimeters
and contains the following requirements:
track is associated with LKr with E/p between 0.96 and 1.03,
there are no hits in MUV3 associated with the track.

The {\bf kinematics} of the Ke3 decay is used to further clean the sample.
The kaon is identified by a KTAG candidate close in time
with the trigger:
 $|t_{KTAG} - t_{track}| < 1$ ns. 
Each kaon is assigned the average momentum obtained from
a sample of fully reconstructed K$^+\to\pi^+\pi^-\pi^+$
decays, instead of the value measured by the GTK,
since the GTK performance was not optimal in 2016.
The kaon and positron tracks are required to form a vertex 
with  $110 < z < 180$ m and
 $d_{min}<$ 25 mm, where d$_{min}$ is the minimal distance between the tracks.
The neutral pion is reconstructed from two clusters in LKr
 not associated with the track,
with no signal in photon veto detectors (LAV, IRC, SAC).
The missing mass squared, 
assuming the positron hypothesis for the track,
is requested to be close to 0:
 $|P_K - P_e - P_{\pi^0}|^2 <$ 0.01 GeV$^2$/c$^4$.
To reject the residual background from the K2$\pi$ decay,
the missing mass squared, assuming the pion hypothesis, is 
required to be outside the interval (0,~0.04)~GeV$^2$/c$^4$.

Finally, the {\bf RICH selection} is performed to have a sample of single
ring events. The number of hits per ring is 
requested to be greater than three. 
The ring is required to lie within PM acceptance.
A corresponding circle at the mirror plane, constructed
 as explained in section~\ref{alignment_procedure},
is requested to be within the mirror acceptance.
Also, to avoid possible light loss,
the selection contains the requirement for Cherenkov cones
not to have intersection with the beam pipe
(the latter condition is checked for the largest cone
corresponding to the most upstream light emission point).

To precisely measure the ring parameters and correctly calculate the number of hits, 
a standalone {\bf iterative single ring fit} algorithm
has been developed. At each step 
a standard single ring fit is performed:
the sum $\sum\limits_{i}(r_i - R)^2/\sigma_{hit}^2$ is minimized, where $r_i$ is the 
distance between the i-th hit position and the ring centre, 
R is the ring radius, 
$\sigma_{hit}$=4.7 mm is
the single hit resolution (see section~\ref{single_hit_resolution}).
After the ring fit,
 a special $\chi^2(iter)$ is calculated for each hit: 
 $\chi^2(iter) = (r_i - R)^2/\sigma_{hit}^2 + (t_i - \bar{t})^2/\sigma_t^2$.
Here 
t$_i$ is the i-th hit time, $\bar{t} = \frac{1}{n} \sum\limits_{i} t_i$ is the average hit time,
 $\sigma_t$=0.28~ns is the hit time resolution.
The hit with the largest $\chi^2$(iter) is removed and the ring fit is repeated
 unless one of the following conditions 
is satisfied: 
\begin{itemize}
\item $\chi^2$(iter) <  16 for each hit;
\item N$_{iter} >$5;
\item N$_{hits}$=4. 
\end{itemize}

The iterative procedure allows to effectively remove
noise hits that are far from the main bulk of hits in space and/or time. 
On average, 0.8 hits per event are rejected. 

\subsection{Ring radius resolution}

\begin{figure}[h]
\centering
\includegraphics[width=7.5cm]{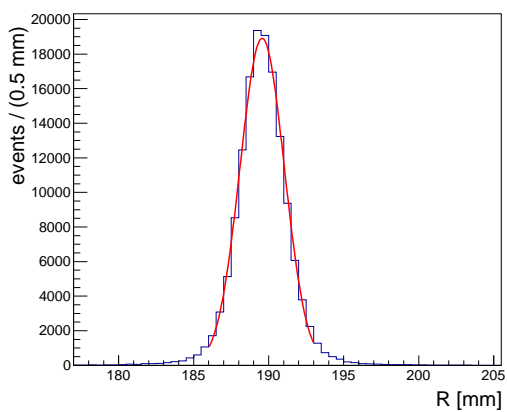}
\caption{Positron ring radius. A gaussian fit is performed: <R>=189.6 mm, $\sigma_R$=1.47 mm.}
\label{radius_resolution}   
\end{figure}

\hspace{15pt}
The ring radius distribution is shown in figure~\ref{radius_resolution}.
The ring resolution is obtained from the gaussian width of the distribution.

\subsection{Ring centre resolution}
\begin{figure}[h]
\begin{minipage}[t]{0.45\textwidth}
\centering
\includegraphics[width=7.5cm , angle=0]{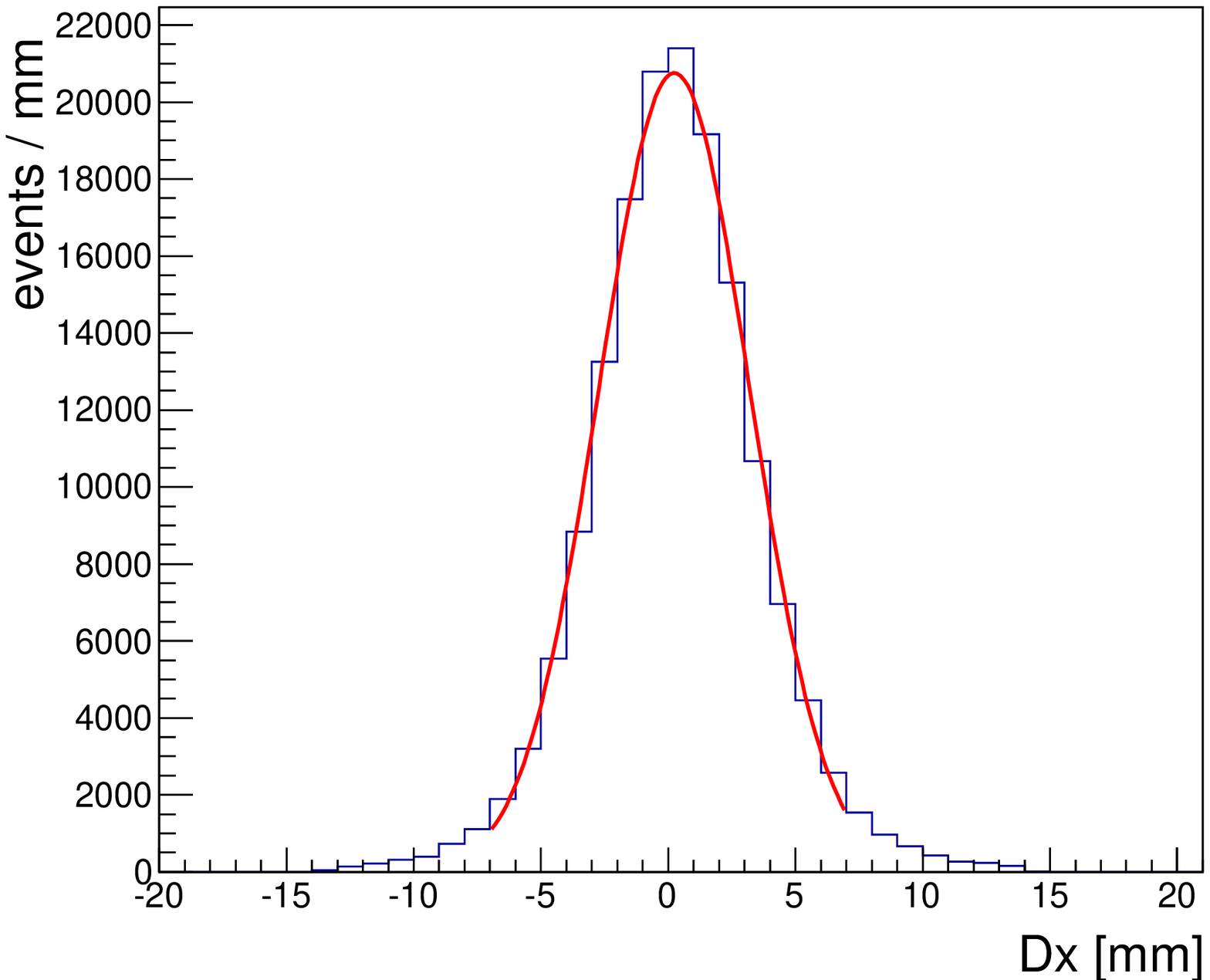}
\end{minipage}  
\hspace{0.2cm}
\begin{minipage}[t]{0.45\textwidth}
\centering
\includegraphics[width=7.5cm , angle=0]{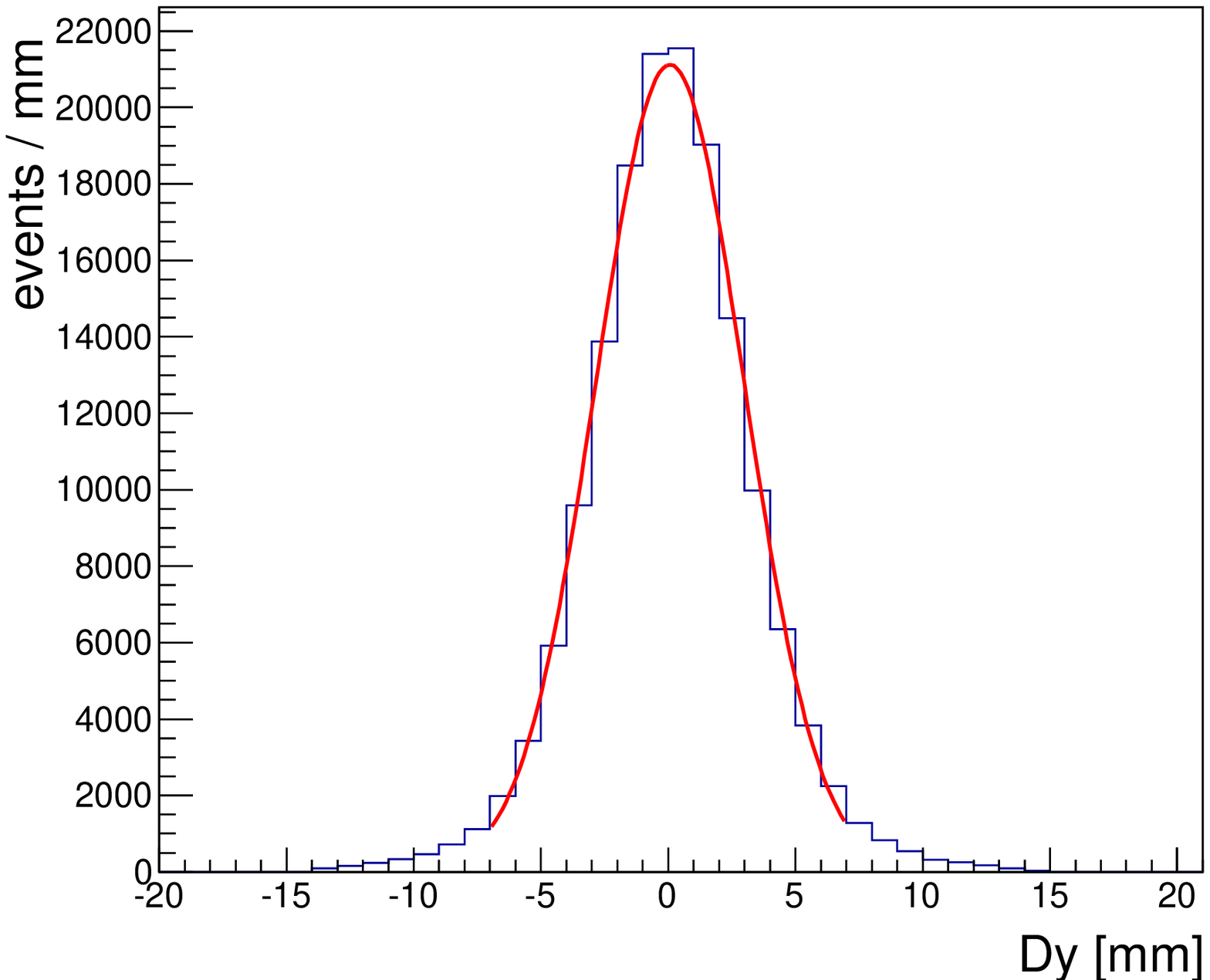}
\end{minipage} 
\caption{Difference between the measured and expected positron ring centre position. 
 A gaussian fit gives $\sigma_x$=2.96 mm (left) and  $\sigma_y$=2.92 mm (right).}\label{ringcentre_resolution}
\begin{picture}(1,1)
\end{picture} 
\end{figure}

\hspace{15pt}
To estimate the ring centre resolution, the 
difference  between the 
measured and expected
ring centre position (in X and Y)
 is plotted and fitted by a gaussian, see figure~\ref{ringcentre_resolution}.
The uncertainty of the expected ring centre position
is determined by the STRAW angular resolution
(see~\ref{alignment_in_2016}) and is much smaller than the measured widths
$\sigma_x \simeq \sigma_y \simeq$ 3~mm, hence
$\sigma_x$ and $\sigma_y$ are used to estimate the ring centre resolution.


\subsection{Single hit resolution}
\label{single_hit_resolution}
\begin{figure}[h]
\centering
\includegraphics[width=7cm]{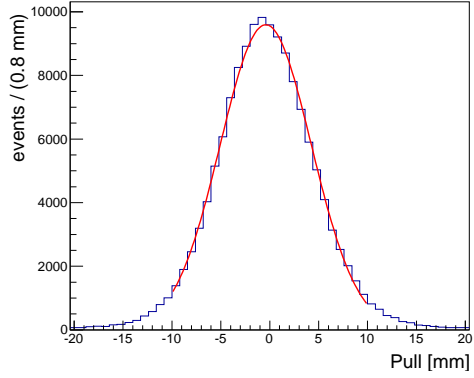}
\caption{Pull distribution. A gaussian fit is performed: $\sigma_{hit}$=4.66 mm.}
\label{singlehit_resolution}   
\end{figure}

\hspace{15pt}
The single hit resolution $\sigma_{hit}$ is estimated from the gaussian 
width of the pull distribution.
The pull is defined as follows: 
Pull = (R - R$_{exp}$)$\sqrt{N_{hits}-3}$.
Here R is the ring radius, R$_{exp}$ is the radius calculated
from the momentum assuming the positron mass, 
(N$_{hits}$-3) is the number
of degrees of freedom of the single ring fit, where
3 is the number of fit parameters (ring radius and two ring centre
coordinates). 
The pull distribution is shown in figure~\ref{singlehit_resolution};
the obtained value is $\sigma_{hit}$=4.66 mm.
 
The main contribution to the single hit resolution 
comes from the geometry, i.e.
from the size of outer and inner Winston cone diameter.
In case of the full light collection by the cone the
geometry contribution is equal to
$\sigma_{geom,~max} = d_{cone}/4 = 4.5$ mm. 
In the opposite case (absorbing cone surface) it is determined by the diameter of the 
sensitive region of PMs: 
$\sigma_{geom,~min} = d_{PM}/4 = 1.9$~mm. 
The mean cone reflectivity
is estimated 
by averaging the Mylar reflectivity over the real photon spectrum.
This spectrum is obtained taking into account all possible effects:
the emission spectrum of Cherenkov photons,
mirror reflectivity,
transmission of quartz windows 
located between cones and PMs,
PM quantum efficiency.
A simple simulation of the hit coordinate spread,
taking into account the calculated mean reflectivity and
assuming not more than one reflection per photon
on the cone with the nominal diameter $d_{cone}$,
gives the following estimate:
 $\sigma_{hit,~ideal~geom} \simeq$4.45~mm. 

The second contribution comes from the mirror misalignment and is 
calculated from the quadratic difference between the
single hit resolution measured on a standard and
``single mirror `` selection (see 
section~\ref{residual_misalignment_contribution}): 
$\sigma_{hit,~mirror}$=2.1 mm.

The contribution due to the
neon dispersion~\cite{neon_dispersion}
can be calculated from the standard deviation $\Delta n$
of (n-1):
$\sigma_{hit,~ \Delta n} \simeq f  \Delta \theta_n \simeq f \Delta n /\theta$,
where $\theta$ is the Cherenkov angle,
$\theta \simeq R/f$. 
The value of $\Delta n = \sqrt{<(n-1)^2> - <(n-1)>^2}$ is obtained
by averaging (n-1) and (n-1)$^2$ over 
the real photon spectrum.
With $\Delta n \simeq 0.4\times 10^{-6}$, this results in
$\sigma_{hit,~ \Delta n} \simeq$0.6 mm which is small compared
to other contributions.

By quadratically subtracting $\sigma_{hit,~mirror}$
and $\sigma_{hit, ~\Delta n}$
from the measured value
$\sigma_{hit}$, the real geometry contribution
can be extracted: $\sigma_{hit,~ real~ geom}\simeq$4.14 mm
which is smaller than $\sigma_{hit,~ ideal~ geom}$.
It could be
due to light losses in multiple reflections
of photons that are incident on
the cone periphery, as described in~\cite{winston_cone_paper}.

\subsection{Number of hits per ring and figure of merit}
\begin{figure}[h]
\centering
\includegraphics[width=8cm]{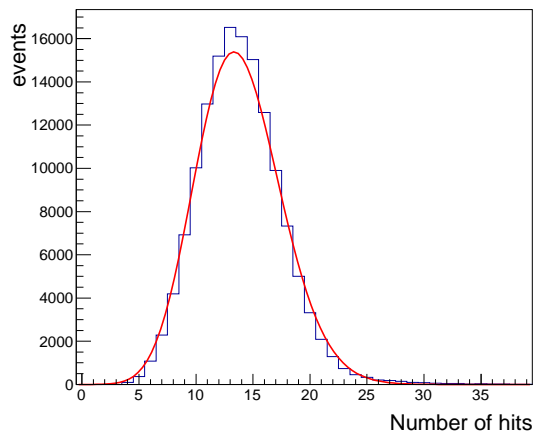}
\caption{Number of hits per ring distribution. A poissonian fit is performed: <N$_{hits}$>=13.8.}
\label{nhits_resolution}   
\end{figure}

\hspace{15pt}
The distribution of the number of hits per ring is shown in figure~\ref{nhits_resolution}.
From the average value of <N$_{hits}$> one can calculate
the figure of merit N$_0$ used to evaluate the performance
of RICH detectors: N$_0 = <N_{hits}>/(L~sin^2\theta)$,
where L is the vessel length and $\theta$ is the Cherenkov angle.
The obtained value is N$_0 \sim$65 cm$^{-1}$.

\subsection{Contribution of the residual mirror misalignment to the performance}
\label{residual_misalignment_contribution}
\hspace{15pt}
To estimate the contribution of the residual misalignment to the
resolutions of ring parameters, the parameter calculation is repeated for the 
events where all the light comes from a single mirror.
The contribution due to the mirror misalignment
is given by the quadratic difference between the
initial and $"$single mirror$"$ value.

Table~\ref{summary_table} summarizes the performance measurements and the contributions due to 
the residual mirror misalignment.

\begin{table}[h]\normalsize
  \centering
  \begin{tabular}{|l|l|l|l|}  
    \hline  
    
    Parameter   &  all events   &   $"$single mirror$"$ events  & misalignment contribution  \\
    \hline
    <R>, mm  &  189.6 & 189.1 & -- \\
    $\sigma_R$, mm  &  1.47 & 1.31 & 0.7 \\
    $\sigma_x$, mm  &  2.96 & 2.82 & -- \\
    $\sigma_y$, mm  &  2.92 & 2.83 & -- \\
    $\sigma_{hit}$, mm  &  4.66 & 4.18 & 2.1 \\
    <N$_{hits}$>  &  13.8 & 14.1 & -- \\
    
    \hline
  \end{tabular}  
  \caption{Performance summary.} \label{summary_table}
\end{table}

A higher <N$_{hits}$> value for $"$single mirror$"$ events
is due to the fact that in this case the mirror  edges with worse reflectivity
are not illuminated.

\section{Conclusion}
\hspace{15pt}
The procedure of the precise RICH mirror alignment has been developed and successfully
accomplished in 2016.
The achieved residual misalignment is $\sim$1~mm in terms of the ring centre position
($\sim$30 $\mu$rad in terms of the mirror angular orientation).

The basic performance parameters have been measured for positron tracks. The ring radius resolution is 1.5 mm, 
the ring centre resolution is 3.0 (2.9) mm for X (Y) coordinate,
the single
hit resolution is 4.7 mm, the average number of hits per ring is 13.8. The contribution of the residual
mirror misalignment to the single hit resolution is 2.1 mm and less than 1 mm to the ring radius
resolution.

\acknowledgments
\hspace{15pt}
The authors are grateful to the staff of the CERN laboratory and the technical staff of participating
universities and laboratories  for their valuable help during the mirror alignment procedure.
 The present  work was completed thanks to the dedication of the whole NA62 Collaboration in operating the experiment 
in data-taking conditions and later in providing  results from the off-line data processing.

\newpage
\appendix
\section{Residual misalignment of all mirrors}

In this appendix the residual misalignment of all mirrors 
at the end of the alignment procedure
is summarized in a table.

\begin{table}[h]\normalsize
\centering
    \begin{tabular}{|l|l|l|l|l|l|}  
      \hline  
   
 Mirror   &  Group   &   X, mm  & $\delta$X, mm & Y, mm  & $\delta$Y, mm \\
\hline
      1   &  Jura     &  -0.9    &  0.3    &  0.5    &  0.1 \\
      3   &  Jura      &  0.2    &  0.1    &  0.9    &  0.1 \\
      4   &  Jura      &  0.1    &  0.1    &  0.0    &  0.1 \\
      5   &  Saleve    &  0.6    &  0.1   &  -0.8    &  0.1 \\
      6   &  Saleve   &  -0.4    &  0.1   &  -1.0    &  0.1 \\
      8   &  Saleve    &  0.0    &  0.1   &  -0.1    &  0.1 \\
      9   &  Saleve   &  -0.4    &  0.1   &  -0.1    &  0.1 \\
     10   &  Saleve   &  -0.3    &  0.1   &  -0.2    &  0.1 \\
     11   &  Saleve   &  0.0    &  0.1    &  0.1    &  0.1 \\
     12   &  Jura      &  0.2    &  0.1   &  0.0    &  0.1 \\
     13   &  Jura     &  -0.1    &  0.1    &  0.3    &  0.1 \\
     14   &  Saleve    &  0.6    &  0.1    &  0.5    &  0.1 \\
     15   &  Jura      &  0.2    &  0.1    &  0.3    &  0.1 \\
     16   &  Jura      &  0.3    &  0.1    &  0.2    &  0.1 \\
     17   &  Saleve   &  -0.4    &  0.1    &  0.2    &  0.1 \\
     20   &  Saleve    &  0.2    &  0.1   &  -0.1    &  0.1 \\
     21   &  Jura     &  -0.1    &  0.1   &  -0.7    &  0.1 \\
     22   &  Jura      &  0.0    &  0.1    &  0.1    &  0.1 \\
     23   &  Jura     &  -0.1    &  0.1    &  0.4    &  0.1 \\
     24   &  Saleve    &  0.6    &  0.3    &  0.9    &  0.2 \\

	  \hline

       \end{tabular}  
	\caption{Residual mirror misalignment.
X and Y are the residual misalignment values, $\delta$X and $\delta$Y are misalignment errors.}\label{table_misalignment}
\end{table}

\end{document}